\documentstyle[epsf]{article}
\begin{document}
\begin{center}
{\Large
Exact Solutions for Cosmological Perturbations with
Collisionless Matter}
\end{center}
\begin{center}
Dominik J. Schwarz \footnote{e-mail: dschwarz@ecxph.tuwien.ac.at}\\
Institut f\"ur Theoretische Physik, Technische Universit\"at Wien,\\
Wiedner Hauptstra\ss e 8-10/136, A-1040 Wien, Austria
\end{center}

\begin{abstract}
\noindent
All regular and singular 
cosmological perturbations in a
radiation dominated Einstein-de Sitter Universe with collisionless
particles can be found by a generalized power series ansatz.
\end{abstract}

\noindent
Although the study of cosmological perturbations 
has a long history 
\cite{Lifshitz}, exact solutions in the linear regime
are known for a few perfect fluid
models only (see, e.g., \cite{Kodama}). Within a more
fundamental description of matter no such solutions have been 
found until recently 
\cite{Kraemmer,Rebhan92a,Rebhan94}.
In its early epochs the Universe contains various 
collisionless, massless particles. Examples are  
background gravitons generated during the Planck epoch and 
neutrinos after their decoupling from the electrons. 
  
Cosmological perturbations for collisionless matter have been analyzed
numerically (e.g., \cite{Stewart72}) and analytically in
the short- and long-wavelength approximations 
\cite{Zakharov79}. Two kinds of solutions show up: Regular solutions
are specified by bounded metric perturbations at the 
Big Bang singularity, whereas for singular solutions 
the metric blows up,
changing the type
of the initial singularity. The latter show superhorizon 
oscillations as discussed by Zakharov 
\cite{Zakharov79} and Vishniac \cite{Vishniac82}. 
The general
solution is a combination of both types. Only the regular 
part of the solution
depends on the distribution of the collisionless  particles
at the singularity.

We obtain exact solutions 
for scalar perturbations by a 
generalized power series ansatz. The matter under investigation is
a mixture of collisionless particles and a perfect fluid in a 
radiation dominated Einstein-de Sitter background.
The corresponding discussion
for the vector and tensor perturbations can be found in 
\cite{Rebhan92a,Rebhan94}.
The equations of motion 
are formulated gauge invariantly. The matter perturbations
can either be obtained from the Boltzmann equation 
\cite{Stewart72,Kasai}
or from the finite-temperature graviton
self-energy \cite{Rebhan91}. Reference \cite{Rebhan94} compares the
two approaches.

We use Bardeen's gauge invariant 
metric potentials \cite{Bardeen}. 
This is useful since
the perturbation
of the energy-momentum tensor
\begin{equation}
\label{1}
\delta T^{\mu \nu} = \delta \left( \frac{2}{\sqrt{-g}}
\frac{\delta \Gamma^M}{\delta g_{\mu\nu}} \right)
\end{equation}
is gauge invariant from the definition of the effective action
$\Gamma^M$ (for background-covariant gauges). 
The scalar metric potentials are denoted by $\Phi$ and $\Pi$. They
are related to the gauge invariant density contrast
$\epsilon_m$ (which is 
the density contrast $\delta$ in the local rest-frame of matter)
and the anisotropic pressure $\pi_T$ by
the Einstein equations:
\begin{eqnarray}
\label{2}
{x^2\over 3}\Phi &=& \epsilon_m \ ,\\
\label{3}
x^2 \Pi &=& \pi_T \ .
\end{eqnarray}
The variable $x$ is the
product of the conformal time and the (fixed) 
wavenumber of the perturbation.

The high-temperature limit of the graviton 
self-energy (the second
varation of $\Gamma^M$ in 
Eq. (\ref{1})) for temperatures below
the Planck scale and the kinetic theory yield
the same equations of motion \cite{Rebhan94} for the
metric potentials. 
They form a closed set of equations  
allowing isentropic perturbations in the perfect fluid
component only.
With $\alpha$ being the ratio of the energy density in 
collisionless matter
and the total energy density, and $\Phi = \Phi_{cll} + \Phi_{PF}$,
the trace of the Einstein equations reads:
\begin{equation}
\label{4}
\Phi^{\prime\prime} + {4\over x} \Phi^{\prime} +
{1\over 3}\Phi + {2\over x} \Pi^{\prime} - {2\over 3} \Pi
= 0 \ .
\end{equation}
Equations (\ref{2}) and (\ref{3}) lead to 
\begin{eqnarray}
\label{5}
{x^2 \over 3} \Phi_{\rm cll} &=& - 2 \alpha \biggr[ - \Phi - 2 \Pi +
2 \int_{0}^{x} d x^{\prime} \left(j_0(x-x^{\prime})+\frac3x
j_1(x-x^{\prime}) \right) (\Phi+\Pi)^{\prime}(x^{\prime})
\nonumber \\
&& + 2 \sum_{n=0}^{\infty} \gamma_n (j_0+\frac3x
j_1)^{(n)}(x) \biggr] 
\end{eqnarray}
and
\begin{equation}
\label{6}
x^2 \Pi = - 12 \alpha \biggr[ \int_{0}^x dx^{\prime} j_2(x-x^{\prime})
(\Phi+\Pi)^{\prime}(x^{\prime}) + \sum_{n=0}^{\infty}
\gamma_n j_2^{(n)}(x) \biggr] \ .
\end{equation}
The $\gamma_n$'s are related to the $n$-th momenta 
of the collisionless particles' initial
distribution function \cite{Rebhan94}.
The evolution starts
at the Big Bang (x = 0).

To solve the equations (\ref{4}) -- (\ref{6}) 
we make the ansatz
\begin{equation}
F(x)=C_1 F_{\rm reg}(x) + C_2 x^\sigma F_{\rm sing}(x) \ ,
\end{equation}
where $C_1$ and $C_2$ are arbitrary constants and
\begin{equation}
F_{\rm reg,sing}(x)=\sum_{n=0}^\infty c^{\rm reg,sing}_n 
{x^n\over n!} \ .
\end{equation}
The singular solutions do not depend on the initial conditions
$\gamma_n$. They can be added to any regular solution 
specified by the initial conditions. The exponent
\begin{equation}
\sigma_\pm=-{5\over2}
 \pm\sqrt{5-32\alpha\over20}
\end{equation}
takes complex values for $\alpha > \alpha_{crit} = 5/32$. Thus
superhorizon size oscillations occur. They can not
be taken seriously up to the initial singularity, because nonlinear
effects of the Einstein equation will be important. But they
are necessary to match a mixmaster scenario \cite{Misner} or 
quantum generated perturbations
from a preceding inflationary epoch \cite{Gri}. 
A strictly Friedmannian singularity
rules out the singular solutions.

In the Figs.~1 and 2 we consider a model with three massless neutrino
species and a perfect fluid consisting of relativistic electrons and
photons in equilibrium. For such a model $\alpha \sim 1/2$ 
\cite{Zeldovich}. 

Fig.~1 shows the regular solutions for $\Pi(0) = 0$ and 
$\Phi(0) = 1, \Phi^\prime (0) = 0$, i.e., a perfect fluid at the
origin. The density contrast grows on superhorizon scales and 
oscillates
after horizon crossing ($x/\pi = 1$). On subhorizont scales the
perfect fluid component has constant amplitude, whereas
the collisionless
component decays ($ \sim 1/x$) due to directional dispersion.
\begin{figure}
\vbox to 8cm
{\vss 
\hbox to \hsize 
{\hss
\epsfysize=11cm
\epsfbox{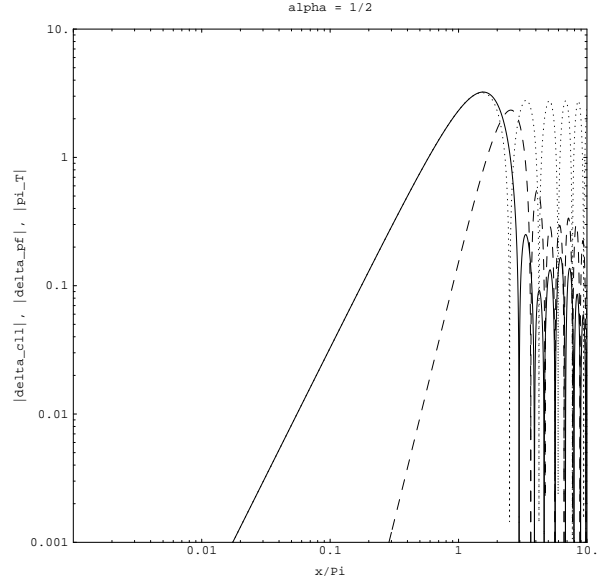}
\hss}
\vss}
\caption{Regular scalar perturbations in a two component universe 
($\alpha = 1/2$). The various lines show regular solutions for
$|\delta_{cll}|$ (full line), $|\delta_{PF}|$ (dotted line), 
and $|\pi_T|$ (dashed line). The initial conditions are specified
by $\gamma_0 = 5/64$, $\gamma_2 = 42 \gamma_0$, and  
$\gamma_4 = 49 \gamma_0$. All other $\gamma_n$'s vanish. }
\end{figure}
\begin{figure}
\vbox to 8 cm
{\vss
\hbox to \hsize
{\hss
\epsfysize=11cm
\epsfbox{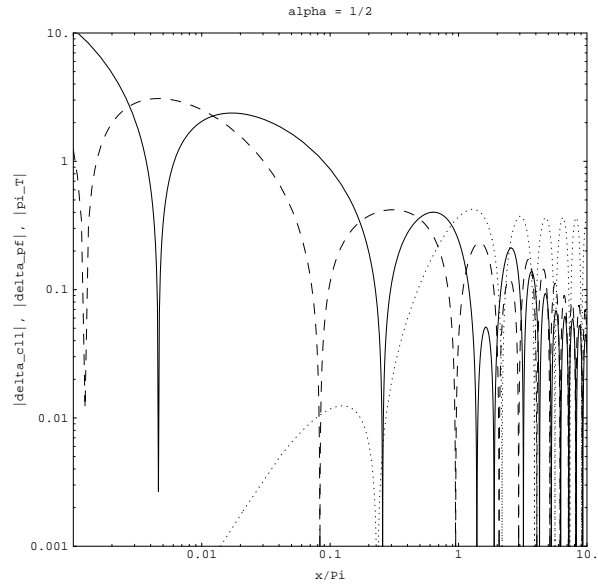}
\hss}
\vss}
\caption{Singular solutions corresponding to $\sigma_+$. 
Same variables as in Fig.~1.
Notice the logarithmic scale in x --- the solutions are
essentially singular as $x \to 0$. The second
solution $\sigma_-$ differs 
by a phase.}
\end{figure}
\clearpage

In Fig.~2 the singular solutions are plotted. The perfect
fluid has the same behaviour
as in Fig.~1, but the collisionless component shows 
superhorizon oscillations as mentioned above. These oscillations
are of geometrical, not of causal, origin. 
Lukash et al. showed that collisionless matter can isotropize
an initial anisotropy \cite{Lukash}. 
Reversing time these solutions reflect the instability
of an isotropic collapse. 
The sound velocity inside the horizon 
equals to the speed of light for all solutions. 

I am very grateful to the organizers for a fellowship to attend the
conference. I thank A. K. Rebhan for a most enjoyable collaboration.
This work was supported by the Austrian ``Fonds zur F\"orderung der
wissenschaftlichen Forschung'' under contract no. P9005-PHY.


\begin{thebibliography}{99}

\bibitem{Lifshitz}E. Lifshitz, Zh. Eksp. Teor. Fiz. {\bf 16}, 587 (1946);
         E. Lifshitz and I. Khalatnikov, Adv. Phys. {\bf 12}, 185 (1963).
\bibitem{Kodama}H. Kodama and M. Sasaki, Prog. Theor. Phys. Suppl. 
         {\bf 78}, 1 (1984);
         V. F. Mukhanov, H. A. Feldman, and R. H. Brandenberger,
         Phys. Rep. {\bf 215}, 203 (1992).
\bibitem{Kraemmer}U. Kraemmer and A. Rebhan, Phys. Rev. Lett. {\bf 67},
         793 (1991); D. J. Schwarz, Int. J. Mod. Phys. D 
         {\bf 3}, 265 (1994). 
\bibitem{Rebhan92a}A. Rebhan, Nucl. Phys. {\bf B369}, 479 (1992);
	in {\it Relativistic Astrophysics and Cosmology}, S. Gottl\"ober,
	J. P. M\"ucket, and V. M\"uller (eds.), (World Sci., Singapore, 1992)
        p. 137; A. Rebhan, Astrophys. J. {\bf392}, 385 (1992).
\bibitem{Rebhan94}A. K. Rebhan and D. J. Schwarz, preprint gr-qc/9403032,
        Phys. Rev. D in press.
\bibitem{Rebhan91}A. Rebhan, Nucl. Phys. {\bf B351}, 706 (1991).
\bibitem{Stewart72}J. M. Stewart, Astrophys. J. {\bf 176}, 323 (1972);
                   P. J. E. Peebles, Astrophys. J. {\bf 180}, 1 (1973);
                   J. R. Bond and A. S. Szalay, Astrophys. J. {\bf 274}, 443 (1983);
                   R. Durrer, Astron. Astrophys. {\bf 208}, 1 (1989);
                   R. K. Schaefer, Int. J. Mod. Phys. A {\bf 6},
                   2075 (1991);
                   C. P. Ma and E. Bertschinger, preprint Caltech GRP-375 (1994). 
\bibitem{Zakharov79}A. V. Zakharov, Zh. Eksp. Teor. Fiz. {\bf 77}, 434 (1979).
\bibitem{Vishniac82}E. T. Vishniac, Astrophys. J. {\bf 257}, 456 (1982). 
\bibitem{Kasai}M. Kasai and K. Tomita, Phys. Rev. D {\bf 33}, 1576 (1986).
\bibitem{Bardeen}J. M. Bardeen, Phys. Rev. D {\bf 22}, 1882 (1980).
\bibitem{Misner}C. W. Misner, Phys. Rev. Lett. {\bf 22}, 1071 (1969);
	V. A. Belinskii, I. M. Khalatnikov and E. M. Lifshitz,
	Adv. Phys. {\bf 19}, 525 (1970).
\bibitem{Gri}L. P. Grishchuk, Sov. Phys. - JETP {\bf 40}, 409 (1975);
	Phys. Rev. D {\bf 48}, 5581 (1993); preprint gr-qc/9405059 
        (1994).
\bibitem{Zeldovich}Y. B. Zeldovich and I. D. Novikov, {\it
        Relativistic Astrophysics Vol. 2: The Structure and
        Evolution of the Universe}, (The Univesity of Chicago Press, 
        Chicago, 1983).
\bibitem{Lukash}V. N. Lukash, I. D. Novikov and A. A. Starobinskii,
        Sov. Phys. - JETP {\bf 42}, 757 (1976).
\end{thebibliography}
\end{document}